**MFC-based biosensor for domestic wastewater COD assessment in constructed wetlands**

**Corbella[1], C., Hartl[1], M. Fernandez-gatell[1], M. and Puigagut[1]\*, J.**

[1]Group of environmental engineering and microbiology (GEMMA), Universitat Politècnica de Catalunya - BarcelonaTech

*corresponding author: jaume.puigagut@upc.edu

Postal address:

C/Jordi Girona 1-3, Building D1-105

08034 Barcelona (Spain)

Tel. +34 934015952

**Abstract**

In the context of natural-based wastewater treatment technologies (such as constructed wetlands - CW) the use of a low-cost, continuous-like biosensor tool for the assessment of operational conditions is of key importance for plant management optimization. The objective of the present study was to assess the potential use of constructed wetland microbial fuel cells (CW-MFC) as a domestic wastewater COD assessment tool. For the purpose of this work four lab-scale CW-MFCs were set up and fed with pre-settled domestic wastewater at different COD concentrations. Under laboratory conditions two different anodic materials were tested (graphite rods and gravel). Furthermore, a pilot-plant based experiment was also conducted to confirm the findings previously recorded for lab-scale experiments. Results showed that in spite of the low coulombic efficiencies recorded, either gravel or graphite-based anodes were suitable for the purposes of domestic wastewater COD assessment. Significant linear relationships could be established between inlet COD concentrations and CW-MFC $E_{cell}$ whenever contact time was above 10 hours. Results also showed that the accuracy of the CW-MFC was greatly compromised after several weeks of operation. Pilot experiments showed that CW-MFC presents a good bio-indication response between week 3 and 7 of operation (equivalent to an accumulated organic loading between 100 and 200 g COD/m$^2$, respectively). Main conclusion of this work is that of CW-MFC could be used as an "alarm-tool" for qualitative continuous influent water quality assessment rather than a precise COD assessment tool due to a loss of precision after several weeks of operation.

## 1.-Introduction

Organic matter content is one of the legally regulated parameters in wastewater treatment processes. Concentration of organic matter compounds in urban wastewater discharges is limited to 125 mgO$_2$/L of COD by the Spanish legislation (MAGRAMA, 2007). Chemical and Biochemical oxygen demand (COD and BOD, respectively) are the main parameters to determine wastewater organic matter content. Although its precise and rapid quantification is crucial, current methodologies are time consuming, produce chemical compounds that pose a threat to the environment and require qualified personnel (Kumlanghan et al., 2007). As a consequence they are not suitable for real-time monitoring thus preventing a rapid response to contamination events (Di Lorenzo et al., 2009; M. Kim et al., 2003). Microbial Fuel Cells are bioelectrochemical systems that generate electricity from the oxidation of organic compounds by means of exoelectrogenic bacteria as catalysts (Logan et al., 2006). Besides the use of MFCs for green energy production, they have the potential to be used as a tool for continuous or semi-continuous organic matter concentration assessment. MFCs directly provide an electrical signal that can be correlated to organic matter content (Peixoto et al., 2011). The main advantages resulting from the utilization of MFCs as biosensor devices are the possibility of in-situ implementation, on-line monitoring and avoiding complex laboratory procedures requiring addition of chemicals and the fact that no transducer is needed (Di Lorenzo et al., 2009; Peixoto et al., 2011). So far several authors have already proved the feasibility of using this technology for biosensing purposes for both synthetic and real wastewater COD assessment (Chang et al., 2004; Di Lorenzo et al., 2009). However, there is a current lack of research on the use of inexpensive electrode materials, simple MFC design and operation conditions that would match the requirements of a MFC-based biosensor tool implemented in CW.

In the context of natural-based wastewater treatment technologies the use of a biosensor tool for the assessment of operational conditions is of key importance for plant management optimization. CWs are natural wastewater treatment systems in which organic matter is removed by means of physical, chemical and biological processes (García et al., 2010). These removal processes occur within the CWs' treatment bed which is generally filled up with gravel and provides a surface for biofilms to establish. Low energy requirements and straightforward operation and maintenance are some of the advantages of CW's (Puigagut et al., 2007). However MFCs constitute a technology based on high cost materials such as the conductive electrode materials or the cation exchange membranes (Logan, 2008). Therefore, the introduction of a biosensing tool in the context of constructed wetlands needs to address the utilization of both inexpensive and robust materials. In accordance with that, gravel has been demonstrated to be a suitable surface for exeoelectrogens to establish (Corbella et al., 2015) as well as a suitable anodic material when an electron collector (such as stainless steel mesh) is provided (Corbella et al., 2016). Therefore, the objective of the present study was to assess the use of a novel membrane-less constructed wetland microbial fuel cell (CW-MFC) as a suitable wastewater COD biosensor for its application in the CW domain. Both gravel and graphite were tested as anodic materials. Furthermore, a follow-up experiment using pilot-scale CW-MFC systems was also conducted and the results obtained are discussed in line with those obtained from lab-scale experiments.

## 2.-Materials and methods

### *2.1.-CW-MFC configuration*

CW-MFCs configuration is shown in Figure 1. Total system volume was 0.85 L. The cylindrical anodic chamber (Ø=9 cm, height=15 cm high) was made of transparent acrylic plexiglas and contained three separate layers of the anodic material, each wrapped with stainless steel mesh (marine grade 316L), that served as the electron collector. The three layers of anodic material were externally connected by means of stainless steel wires resulting in a single anodic electrode. Two different anodic materials were tested in the experiment: gravel (GRAV-MFC) and graphite rods (GRAPH-MFC). Four replicates of gravel based MFCs and two replicates of graphite based MFCs were operated. The anodic volume of graphite CW-MFCs was filled with 1 cm long and 6.15 mm wide graphite rods (Alfa Aesar, 99.9995%, metal basis, ACKSP grade, Ultra "F" Purity). Gravel CW-MFCs were filled with granitic gravel commonly used in HSSF CW ($D_{60}$=7.3; Cu=0.8; porosity=40%). The anodic liquid volume was 0.5 L regardless the anodic material. To guarantee a good homogenization within the anodic chamber, wastewater was constantly recirculated by means of a pumping system (Damova MP-3035-6M; Toshiba VF-nC3). A layer of glass wool was placed over the anodic material to avoid any oxygen leaking from the cathode area. The cylindrical cathodic chamber was made of PVC (Ø=19 cm, height=5 cm). The cathode was made using 4 pieces of 60.8 $cm^2$ graphite felt (Alfa Aesar, 1.12 cm thick, 99.9 % carbon purity; metal basis). As indicated in Figure 1, graphite felt pieces were placed inside the upper cylinder occupying its entire surface, and were connected in between them and with the external circuit by stainless steel wires. The graphite felt was kept semi-submerged thus in contact with both liquid media and air (air cathode). The anode and the cathode were externally connected by means of copper wires through an external resistance of 1000 Ω. The generated voltages were measured and stored every 5 minutes across the external resistance by means of a datalogger (Campbell Scientific CR1000).

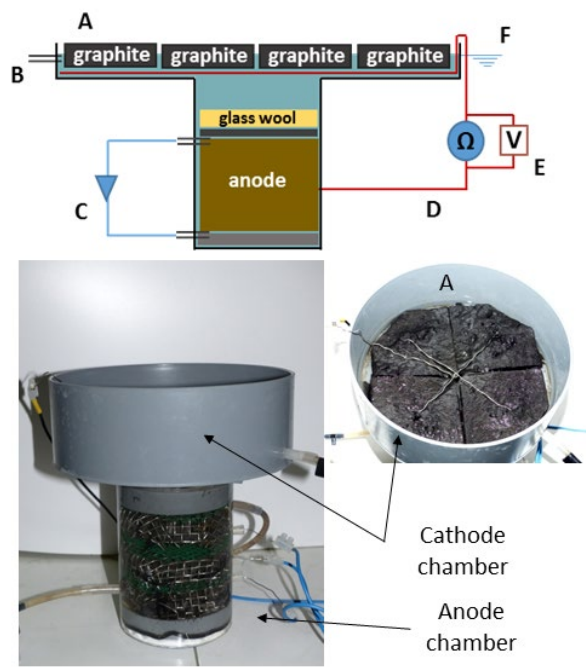

**Figure 1.** Scheme and pics of the CW-MFC biosensor. A: graphite felt cathode; B: Cathode overflow; C: Anodic recirculation; D: External circuit; E: datalogger and external resistance; F: Water level.

## 2.2.-Cathode limitation and start-up

The main objective of this study was to investigate the CW-MFC bioindication capacity and aimed at assessing organic matter oxidation processes occurring within the anodic chamber. In order to ensure that cathodic reactions were not limiting the system, i.e. the cathode could cover the anodic current demand, a preliminary experiment was carried out to determine the optimal cathode to anode surface ratio (C:A). Previous experiments of our group determined that the optimal anode to cathode ratio was 4:1 using graphite as the anodic material (Corbella et al., 2015). In this case, the experiment was conducted with two MFCs with an anode based on gravel, a non-conductive anodic material. The cathode limitation experiment was conducted by step-wise increasing the cathodic area resulting in the following C:A ratios: 0.8, 1.6, 2.4, 3.2, 4 $cm^2_{cathode}/cm^2_{anode}$. The cathodic area was increased once the voltage generated by the previous condition was stable for 3h. Voltage was recorded every 5 minutes. After the optimal cathode to anode ratio was determined, cells were cleaned and daily fed with urban wastewater (338±110 mg $COD_{tot}$/L) for 7 days until the experiment of the abiotic reactions described below was performed.

## 2.3.-Assessment of current generation by abiotic reactions

The organic matter bioindication capacity relies on the ability of electrogens to oxidize organic compounds and generate electricity. Bioindication accuracy (considered to be the level of correlation between the COD and the $E_{cell}$ of the systems) can be affected by the generation of current stemming from the direct anodic oxidation of non-organic substrates through abiotic reactions. To quantify the electrons generated through abiotic reactions at the anode, an additional experiment was carried out. This experiment was performed under stable bioelectrochemical conditions. The experiment consisted of comparing the electricity generated by biotic and abiotic MFCs. To this aim, two graphite based cells and two of the gravel-based cells were cleaned (intensely with tap water and with a solution of 1N NaOH following the procedure described in Reguera et al. (2011) and sterilized (autoclave, 20 min, 121 ºC, 1 bar) (ABIO_MFCs). The other two gravel-based cells were kept unaltered (BIO_MFCs). The 6 systems (gravel BIO_MFCs; gravel and graphite ABIO_MFCs) were fed every 48 hours with sterilized wastewater (autoclave, 20 min, 121 ºC, 1 bar) to guarantee an abiotic environment in ABIO_MFCs. In between feeding events, ABIO_MFCs were cleaned by soaking the anode material in a solution of 1N NaOH thus removing the organic matter and biofilm adhered (Reguera et al., 2011). The generated current was recorded every 5 minutes by means of a datalogger (Campbell Scientific CR1000) during the whole experiment.

### 2.4.-Bioindication assessment

The bioindication capacity was tested by feeding the systems with domestic wastewater. Wastewater was collected from the pilot plant described in Corbella and Puigagut (2014), more precisely from the outlet of an anaerobic primary treatment (hydrolytic upflow sludge blanket reactor, HUSB) which pre-hydrolyses the organic matter present in wastewater (Ligero et al., 2001). After its collection, wastewater COD was analyzed and diluted with tap water in order to achieve concentrations of around 25, 50, 75, 100, 150 and 200 mg COD/L. Only for the fourth batch, COD tested concentrations covered a wider range (50, 75, 100,150, 200, 250, 300, 350, 400 and 450 mg COD/L). Tested COD concentrations are in the range of influent/effluent concentrations reported for full-scale HSSF CWs (Puigagut et al., 2007). Once diluted, wastewater was frozen until it was used in order to prevent any biological activity and consequent degradation of the organic matter.

CW-MFCs (2 graphite and 2 gravel replicates) were fed with increasing COD concentrations every 24 hours and the $E_{cell}$ was recorded using a datalogger (Campbell Scientific CR1000). The COD concentration at the beginning of each batch was 0 mg/L (only tap water - control); this was done in order to analyze the effect of the previously retained solids within the system on the electrical signal. Wastewater for each concentration was taken out from the freezer the night before its utilization. Samples were taken from the inlet wastewater and from the anodic chamber after ca. 24 hours of feeding. The experiment was replicated 4 times during 1.5 months (E1 to E4).

### 2.5.-Analytical methods, calculations and statistical analyses

Samples taken were stored at 4 ºC before their analysis. Soluble and total COD were analyzed in triplicate according to Standard Methods (APHA, 2005). Conductivity and pH of the inlet wastewater was monitored during the E4 to detect any influence on system's performance (Endress+Hauser CLM381 and CRISON pH/mV 506). Electrical parameters including coulombic efficiencies were calculated on the basis of total COD according to (Logan et al., 2006). The statistical significance of the linear regressions described in the paper were determined using the software package provided in excel 2016 that allows to perform an ANOVA test on linear regressions. Results were considered significant (*) under p-values below 0.05, and very significant (**) under p values below 0.01. The F value, the degrees of freedom (DF) and the critical value for F (p-value) are indicated for each linear regression performed.

### 2.6.-COD bioindication in pilot-scale systems

For the purpose of this experiment, three pilot-scale CW-MFC systems were designed, built up and monitored for ca. two months. The CW-MFCs consisted of a PVC reservoir (length=55 cm, width=35 cm, wetted filling height=28 cm, area=0.193 $m^2$) filled with 4/8 mm granitic riverine gravel. The systems were continuously fed with primary treated domestic wastewater that was stored up to three days under agitation within a reservoir of ca. 180 L of volume. The reservoir was re-filled with fresh primary settled domestic wastewater three times per week, leaving the water between 2-3 days within the reservoir. Organic matter concentration within the influent reservoir decreased along the two- or three-day-period; therefore, organic matter concentration entering the systems was highly variable, allowing to effectively track the MFC's electrical response to the organic matter variation. The systems were operated at a flow rate of 5 L/d, resulting in a theoretical HRT and average OLR of 3.9 days and 4.5 g COD/$m^2$.day, respectively.

The CW-MFCs were designed as three MFC electrodes in series along the flow path (Figure 2). Each electrode consisted of an anode based on four stainless steel meshes (marine grade A316L, mesh width=4.60 mm, Øwire=1.000 mm, S/ISO 9044:1999) in series (each one 4 cm away from each other). Each metal mesh covered nearly the whole cross sectional area (0.08 $m^2$) of the CW. The three cathodes consisted of a carbon felt layer (Alfa Aesar, 1.27 cm thick, with a surface of 0.03 $m^2$, 99.0% carbon purity) which was placed on the surface of the gravel bed and kept semi-submerged (as recommended elsewhere – Corbella et al. 2016a), thus in contact with both liquid media and air (air cathode). Each electrode's anode and cathode was externally connected via a 220 Ω resistance. The voltage across the external resistance for each electrode was continuously monitored by means of a datalogger (Campbell Scientific CR1000, AM16/32B Multiplexor). However, for the purpose of this work, only the recorded cell voltage of the first electrode (the one closest to the influent) will be considered. Water samples were taken from the influent at different time intervals and were processed for total chemical oxygen demand (COD) following Standard Methods (APHA, 2005).

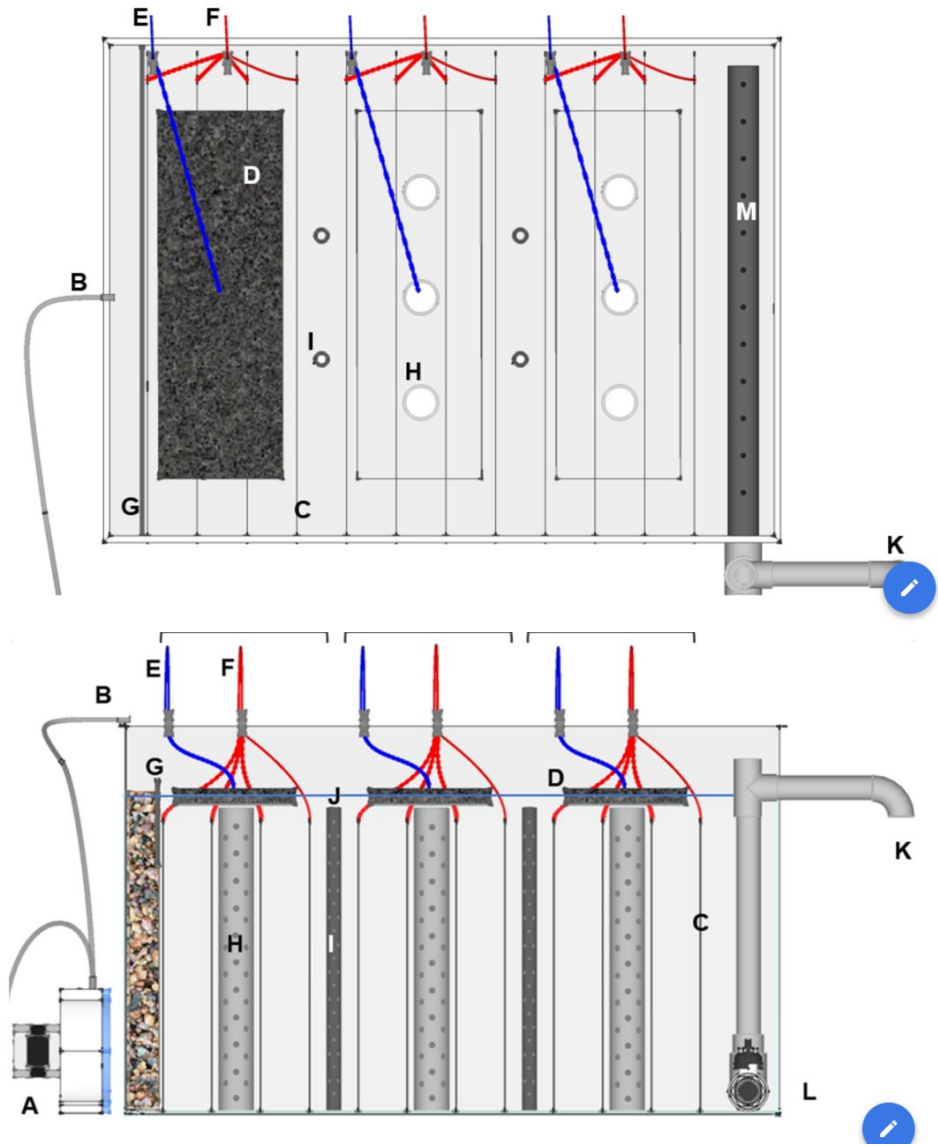

**Figure 2.** Section- (top) and plan-view (bottom) of the CW-MFC systems. A: Pump; B: Inflow; C: Anode; D: Cathode; E/F: Anode/Cathode connection to datalogger; G: Inflow barrier to avoid water short-circuiting on surface; H: Gravel core sampling tubes; I: Liquid sampling tubes; J: Water level; K: Standing pipe effluent; L: Drainage; M: Effluent collection tube.

# 3.-Results

## *3.1.-Abiotic reactions assessment*

Bioindication accuracy depends on to what extend a relationship can be established between the chosen electric parameter (current in our case) and the quantity of organic matter present in the media. In MFCs current is described to mainly derive from the organic matter oxidation conducted by the exoelectrogenic bacteria community. However, although being less efficient, abiotic oxidation reactions can also occur within the anodic chamber thus contributing to the overall current production (Lovley, 2006). These abiotically oxidized compounds can be both organic and inorganic (Lovley, 2006). Therefore, if other non-organic compounds that are not quantified in the COD analysis are oxidized using the anode as the electron acceptor, the reliability of the biosensing capacity can be compromised to some extent. Urban wastewater is a very diverse substrate (Metcalf and Eddy Inc., 1991), containing numerous different compounds, such as sulfides, that could abiotically react with the electrode generating electricity (Lovley, 2006; Ryckelynck et al., 2005). Therefore, we assessed the contribution of non-biological reactions to the overall current production so as to determine whether abiotic reactions could be limiting the precision of the CW-MFC as a COD assessment tool.

Results of this experiment are presented in Figure S1 (supplementary material). As it can be seen, the number of electrons transferred during 48 hours was statistically higher for BIO_MFCs ($p<0.05$). Actually, the electrons transferred in ABIO_MFCs generally accounted for less than 2% of the electrons transferred in biological systems regardless the anodic material considered. Accordingly, it can be concluded that abiotic reactions did not contribute significantly to the electrical current and therefore they could not negatively affect the biosensing accuracy of the MFC.

### 3.2.-Cathode limitation assessment

Voltage generated by MFCs is always diminished by overpotentials and ohmic losses which can occur both in the anodic and the cathodic compartments (Logan et al., 2006). Cathodic overpotentials are main contributors to overall performance losses in MFCs (Rismani-Yazdi et al., 2008) due to the slow kinetics of the oxygen reduction at graphite electrodes (Gil et al., 2003). Since the goal of the presented research was to test the electrical generation as function of the organic matter concentration introduced in the anodic chamber it was crucial to ensure that the cathode was not limiting the system's performance (Chang et al., 2004). One of the strategies to lower cathodic overpotentials is increasing the electrode surface area (Logan et al., 2006). Accordingly, experimental trials were carried out to determine the cathode to anode ratio (C:A) that did not limit the system, thus maximizing the voltage generated. Results are presented in Figure S2 and, as it can be seen, voltage increased concomitantly with the cathodic area until the C/A ratio was that of ca. 3.0 $cm^2/cm^2$. From then on, although the ratio was again increased up to 4 $cm^2/cm^2$, no statistical differences were found between voltages generated (Figure S2 – Supplementary material). Similar results were reported for conventional MFCs fed with acetate as carbon and electron source (Oh and Logan, 2006). Results obtained in this experiment suggested that, when wastewater is used as substrate, a ratio between 3 to 4 $cm^2/cm^2$ is enough to avoid cathodic limitation. According to this result and considering that the optimal ratio for graphite based MFCs had been previously set to be 4:1 (Corbella et al., 2015), the latter was the ratio implemented for the bioindication experiment.

### 3.3.-Conductivity effect

Solution conductivity is also one of the factors contributing to ohmic losses (Logan, 2008) and thus an influencing parameter on MFC voltage generation. Actually, one of the drawbacks of sustainable wastewater treatment using MFCs is the low solution conductivity of wastewaters (Liu and Cheng, 2014). Due to wastewater dilution with tap water, conductivity was altered leading to different conductivities along the experiment. Therefore, during E4 wastewater conductivity was measured in order to assess the effect of solution conductivity on MFC voltage generation. As shown in Figure 5 there was a clear relationship between conductivity and COD (both total and soluble) indicating that wastewater dilution affected the conductivity of the sample. In accordance with this result, Peixoto et al. (2011) also reported a clear relationship between power and conductivity. Furthermore, Peixoto et al. (2011) highlighted that the effect of conductivity was more pronounced between 1.1±0.012 and 2.1±0.012 mS/cm, which matches our experimental conditions.

The positive relationship between both parameters (Figure S3) can question whether MFCs bioindicate solution conductivity rather than OM concentration of the media. Accordingly, there are two different solutions reported in the bibliography to avoid the effect of conductivity in wastewater MFC biosensors. Di Lorenzo et al. (2009) added a phosphate buffer (5M, pH7) to keep the wastewater conductivity constant. Otherwise, Peixoto et al. (2011) suggested the application of correction factors at measurements done under different environmental conditions (temperature, conductivity or pH). However wastewater is a very variable substrate and if biosensor MFCs are to be implemented in wastewater treatment plants, influent wastewater can vary hourly (Metcalf and Eddy Inc., 1991). Consequently, both the addition of phosphate buffers and the application of correction factors would largely increase the complexity of MFC bioindication. Also a positive correlation between conductivity and COD has been reported from the continuous analysis of a WWTP influent wastewater (Daal-rombouts et al., 2013). This positive relationship between the two parameters in real urban wastewater can thus be considered a positive influence regarding bioindication purposes. Therefore, authors decided not to correct the conductivity factor to simulate the real implementation of MFCs as biosensors in CW wastewater treatment plants.

### 3.4.-Voltage patterns and COD removal

Voltages generated after the addition of wastewater within the systems followed a constant pattern along all the experiment both for graphite and gravel MFCs. This representative pattern, which is presented in Figure S4, is in accordance to that described for single-batch operation of conventional MFCs elsewhere (Peixoto et al., 2011). As it is shown, a drastic increase of voltage was generated immediately after the addition of wastewater. Then voltages increased smoothly and continuously to the end of the experiment (after ca. 24h). When the system was emptied, a steep decrease of voltage to zero was recorded. There was no difference between materials in terms of the voltage patterns, however, due to the higher conductivity of the anodic material used, graphite MFCs generated higher voltages and intensities. More precisely, after 20h of contact time GRAPH-MFCs produced intensities of 163±92% higher than GRAVEL-MFCs. However, as it is further discussed, the bioindication capacity of the gravel systems was not compromised because of the lower intensities produced.

Removal efficiencies were calculated at 24 hours of contact time for all the concentrations tested and for the four experiments conducted (E1 to E4). Removal efficiencies increased with inlet COD concentrations (results not shown). Accordingly, regardless the experiment (E1 to E4) and the material considered, removal efficiencies obtained when the highest inlet concentrations were tested (204±52 mg COD/L) averaged 56±9%. These rates are in the lower range of real scale HFCW removal efficiencies (Puigagut et al., 2007).

### 3.5.-Effect of anode material and contact time on Bioindication

In this work, a MFC-based bioindicator tool is proposed to estimate the COD concentration in of domestic. As it is described in this section, the ability of the proposed tool to assess a certain value of COD was not only material dependent, but also depended on the contact time (CT). The CT effect on the performance of the proposed tool was carried out by analyzing the value and the statistical significance of $R^2$ of the linear regression between COD concentration and $E_{cell}$ . Contact times analyzed were 1, 2, 3, 4, 5, 10, 15 and 20 hours after wastewater addition. However, for the sake of simplifying discussion only the results of 1, 5, 10 and 20 hours will be considered. Results showed that the analysis of the relationship between the $E_{cell}$ and the COD concentration was described by a linear equation. The linear relationship between COD and $E_{cell}$ was best after 20h of contact time, regardless the anode material considered (Figure S5 – Supplementary material). This significant linearity was found for both gravel and graphite MFCs. The determined slopes of the linear functions were similar regardless the material considered and ranged from 0.5-1.3 mV/mg COD/L and from 0.6-1.0 mV/mg COD/L to the gravel and graphite MFCs, respectively. Moreover, bioindication was similar across materials tested, although it tended to be better (higher linear regression $R^2$ values) for graphite based MFC (Figure 3 – Figure S6 – Supplementary material). Furthermore, graphite MFC showed the capacity to assess lower values of COD than gravel MFC (Figure 3 – Figure S6 – Supplementary material). That is to say, that at lower COD concentrations, the $E_{cell}$ increase for gravel MFC was lower than for graphite based MFC. According to our results, gravel MFC needed concentrations close to 70 mg/L of COD to produce a significant increase in $E_{cell}$, whereas graphite MFC produced significant increases in $E_{cell}$ from COD concentrations close to 40 mg/L. Therefore, we can conclude that the lower limit of detection is ca. 40 mg COD/L and 70 mg COD/L to the graphite and gravel MFCs, respectively. Several authors used conventional MFCs as biosensing systems and found relationships between inlet organic matter concentrations (either by means of analyzing BOD or COD) and electric output (different electric parameters have been used to calibrate sensors: current, voltage, charge, etc.). The results obtained in this study are in accordance to the bibliography, are statistically significant and provided similar bioindication ranges (significant linear correlations for the tested COD concentrations and the $E_{cell}$) reported (Di Lorenzo et al., 2009; Gonzalez del Campo et al., 2013; B. H. Kim et al., 2003; Peixoto et al., 2011). More precisely, Di Lorenzo et al. (2009), using a conventional air-cathode MFCs, found linear relationships between COD content of both real and synthetic wastewater (in the range of 40 and 200 $mgO_2$/L) and current generated. It is worth mentioning that, to the authors knowledge, this is the first paper that analyses the potential use of a non conductive media (gravel) as electrode material for biosensing purposes during the treatment of real domestic wastewater. The findings here reported are of special interest for plant management in the context of constructed wetlands.

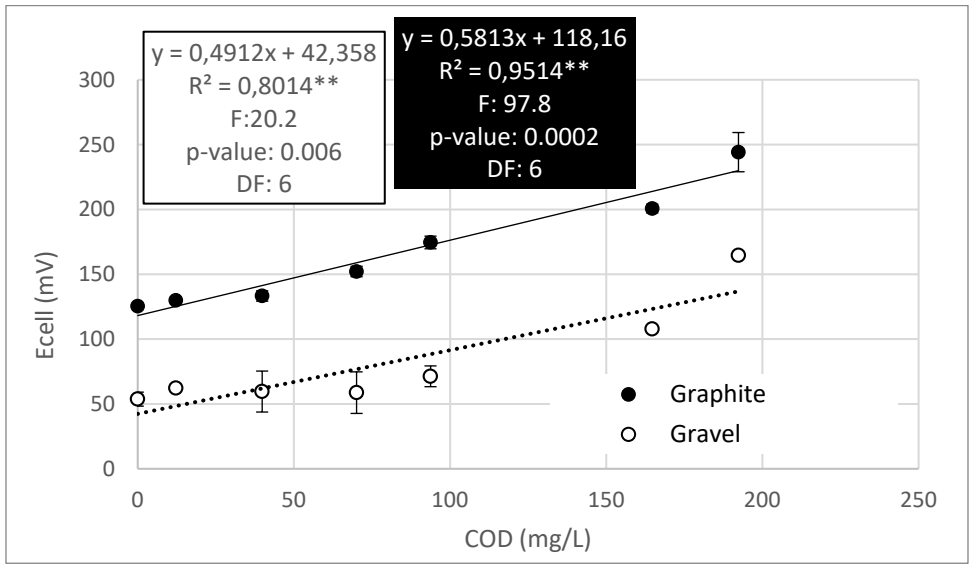

**Figure 3.** Representative Linear relationship established between the $E_{cell}$ and COD concentrations for gravel and graphite CW-MFCs (data from E1 after 20h of contact time). Note: Asterisk represents the level of statistical significance of the linear regression (*p-value<0.05; **p-value<0.01).

Response time is a crucial parameter for biosensing tools, especially for real-time monitoring (Liu and Mattiasson, 2002). Our results showed better R-squared values with increasing contact time (Figure S5). This finding was especially evident for E1 and E2. This trend was observed for both gravel and graphite MFCs. The longer the organic matter remains within the system, the more organic matter is degraded by means of exoelectrogenic pathways thus leading to a more significant relation between the quantity of organic matter and the current generated. Accordingly, our biosensors showed correlations with R-squared values higher than 0.8 for contact times higher than 10 hours regardless the material considered (Figure S5). However, the best results were obtained after 20h of contact time (Figure S5). The contact time appeared to be especially important in case of low organic matter concentrations (Figure 4). While at a contact time of 5h, both gravel and graphite system´s bioindication range (range of detection) was from 95 to 190 $mgO_2$/L, at 20h the lower detection limit decreased to 70 $mgO_2$/L for gravel MFCs and to 40 $mgO_2$/L for graphite MFCs. These results indicate that the contact time had also an influence on MFC's bioindication range. This could be clearly seen in the first experiment in which the lower detection limit decreased with the increment of the contact time (Figure S5). However, it is worth mentioning that graphite systems showed the capacity to bioindicate concentrations over 100 $mgO_2$/L starting from 1h of contact time.

Our results suggest that the time needed to reach a linear relationship between parameters using real wastewater was significantly higher than those reported in literature when using glucose as MFCs carbon source. For example, Chang et al. (2004) reported response times of 60 min when using conventional MFCs fed with glucose and glutamic acid, while Kumlanghan et al. (2007), achieved response times between 3 and 5 minutes when using glucose as the carbon source. However, wastewater is composed by complex carbohydrates that need to be hydrolyzed to volatile fatty acids before being utilized by exoelectrogenic bacteria (Kiely et al., 2011) and therefore, as expected, the response time of systems fed with wastewater are significantly higher. If results obtained in this article are compared to MFC biosensors working with real wastewater they are within the range of results in literature. Peixoto et al. (2011) reports response times of about 10h for concentrations higher than 78±8 $mgO_2$/L. Also Di Lorenzo et al. (2009) reports response times of 13.75h.

However, unexpectedly, if we focus on the performance of the system when operated at low organic matter concentrations and low contact times results show an opposite trend. At low concentrations, higher inlet COD led to lower intensities recorded (Figure 4). This pattern was clearly observed during E3 and E4, regardless the material considered (results not shown). These results suggest a short operational stability of the CW-MFCs, at least when operated under the conditions here considered. As discussed below, the operation of the sensor with real wastewater drastically compromised its long-term stability. Contrary to our results, other studies reported long term stabilities using conventional MFCs. Accordingly, Di Lorenzo et al. (2009) reported 8 months of operation when feeding with artificial wastewater and B. H. Kim et al. (2003), 5 years when feeding wastewater that was collected from a starch processing plant. As discussed in much detail bellow, the authors believe that the unique characteristics of CW as a treatment technology is the cause for this evidence of short-term stability when compared to current literature. In line with this, the accumulation of organics within the gravel bed could be the reason for the short-term stability. This result is in accordance on previous experiences of our group (Corbella et al. 2016).

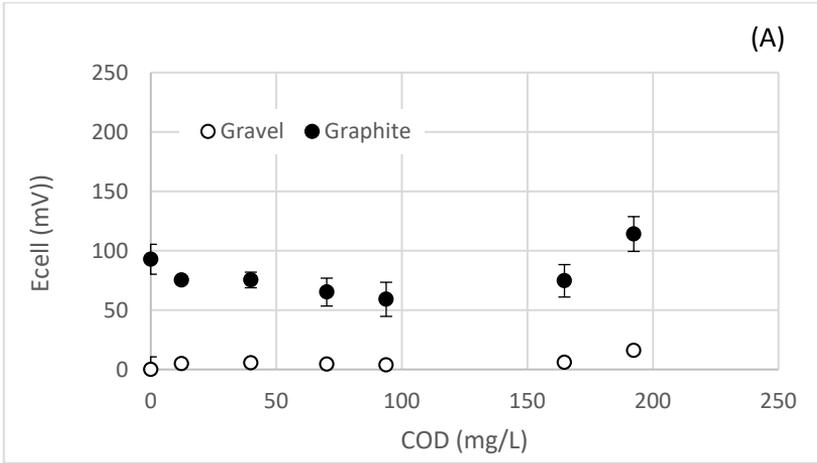

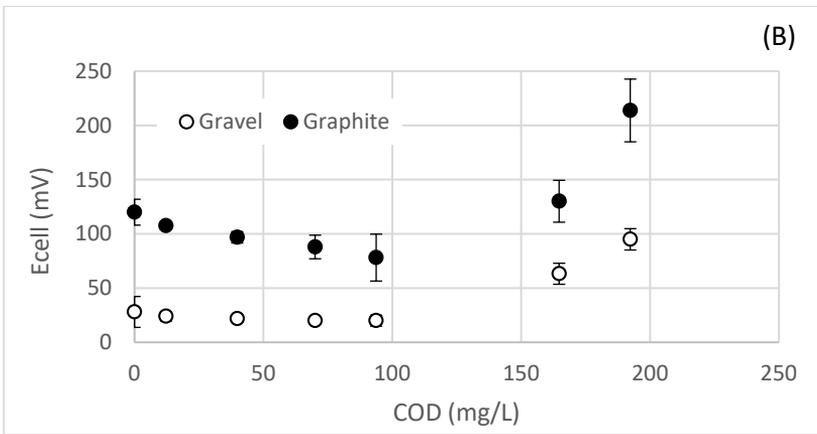

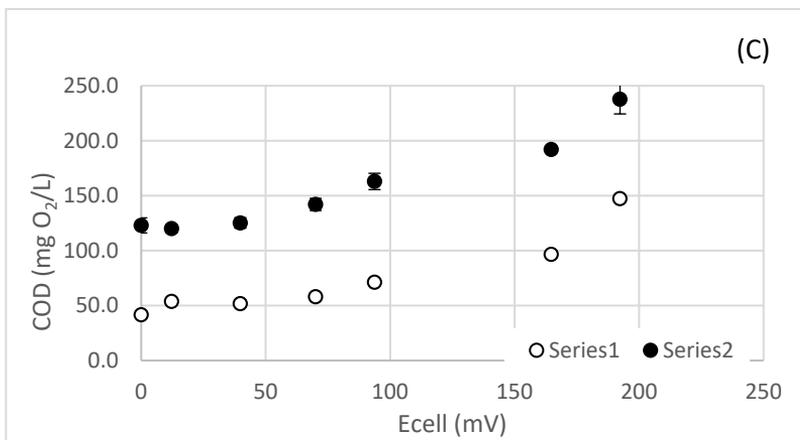

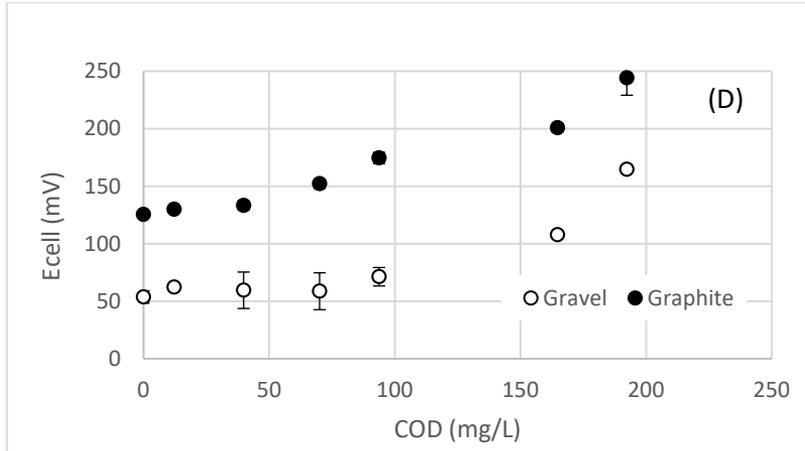

**Figure 4.**Representative figure for $E_{cell}$ plotted against COD concentration tested at different contact times (A):1 hours; (B): 5 hours; (C): 10 hours and D: 20 hours during E1.

### *3.6.-Evolution of the biosensing: effect of the accumulation of solids in the system*

As a consequence of the retention of particulate organic matter, but also of the biofilm generation, the gravel matrix clogs over time in CW (Kadlec and Wallace, 2009) and this is dependent, among other factors, on accumulated organic loading. This organic matter accumulated within the anodic volume can further participate in the generation of electricity and therefore must be determined for the sake of bioindication capacity. It is worth mentioning that here we consider the loss of bioindication capacity in terms of how good was the linear correlation factor ($R^2$) between MFC signal ($E_{cell}$) and COD values. In order to evaluate the effect of organic matter entrapment regarding the bioindication capacity of the tool, at the beginning of each experiment (E1 to E4), systems were fed with tap water (0 mg COD/L) and their electrical response was monitored. As can be seen in Figure 5, both in gravel and graphite MFCs, the $E_{cell}$ generated when tap water was introduced in the system increased with the accumulated organic loading. Therefore, it was demonstrated that the organic matter used by exoelectrogens to generate electricity did not only come from the inlet wastewater, but also from the organic matter already present within the anodic volume. Also endogenous respiration could contribute to the generation of electricity during those periods (Chang et al., 2004). However, after the initial increase, a plateau was reached indicating that there was probably a system limitation (Figure 5). To this regard, clog matter composition in HSSF CWs has been studied and determined to be both of organic and inorganic nature (Knowles et al., 2011). Pedescoll et al. (2013) reported the ratio of volatile solids (VS) with respect to TS to be always less than 50% regardless the operational conditions considered and Caselles-Osorio et al. (2007) determined VS/TS ratios between 10 and 20%. Therefore, the accumulation of inorganic solids within the filter media could lead to a limitation of the transfer rates to the electrode to the point that the electrical response also decreases (M. Kim et al., 2003). At that point, exoelectrogenic communities may have difficulties accessing the substrate due to the accumulation of non-biodegradable solids at the vicinity of the electrode, and may start a process of cell decay near the electrode leading to the decrease in the electricity generated. As a consequence, the MFC signal is worse correlated with COD. . This hypothesis is consistent with the fact that the bioindication capacity in later experiments (E3 and E4) was worse than in former experiments (E1 and E2) (correlation factors between MFC signal provided for a similar COD range decreased to a notable extent along the experiment) (Figure S6). Results corresponding to this decrease in performance as function of organics present in the system are in accordance to those presented by our group in a previous work dealing with the assessment of clogging in constructed wetlands using the electric signal of MFC (Corbella et al. 2016)..

As it can be seen in Figure 5, the MFC signal ($E_{cell}$) response was highly affected by the cumulative organic load, which is in accordance to that of described elsewhere (Gonzalez del Campo et al., 2013; M. Kim et al., 2003). It is difficult to say how long will last the stability of the bioindication tool until recalibration is required, since it is dependent on the accumulated organic loading. Therefore, different systems operating at different organic loading conditions will lead to different required times for recalibration. However, according to our results, the signal of the MFC started to decrease after ca. 500 g COD/m$^2$(Figure 5). Therefore, and in light of results shown in Figure 5, we suggest to recalibrate the system after about 500 g COD/m$^2$. This means, that if we assume standard operation conditions intensity.

for horizontal subsurface-flow constructed wetlands (about 10 g COD/m$^2$.day), the system should be calibrated after about 50 days of operation.

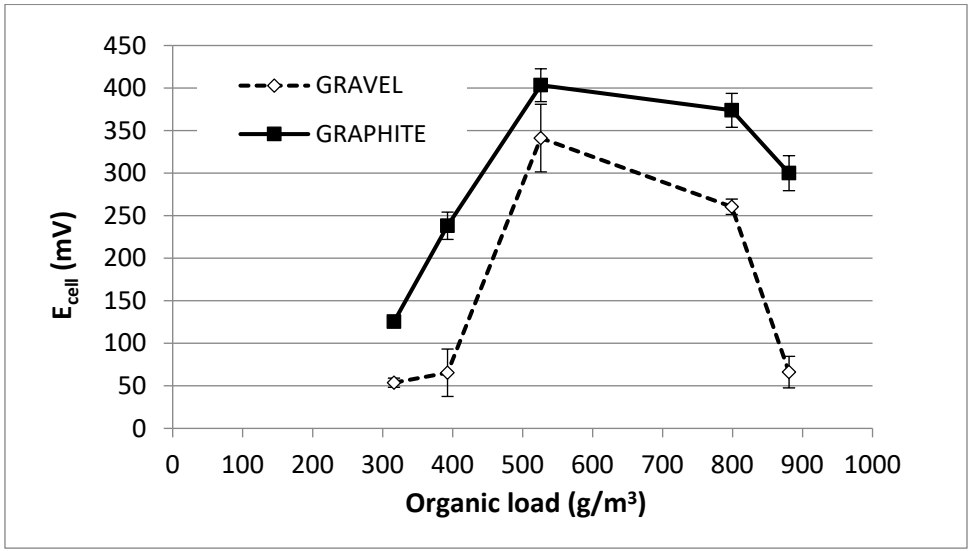

**Figure 5.** $E_{cell}$ recorded under tap water feeding along the experimental period at 20h of contact time.

### 3.7.-COD assessment in pilot-scale systems

Figure 6 shows a representative time lapse for one of the three pilot-scale systems.

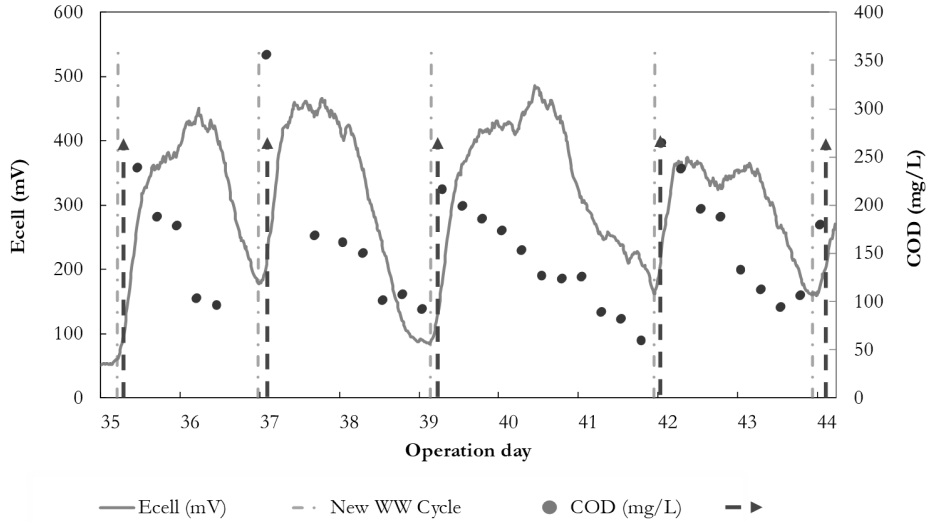

**Figure 6.** $E_{cell}$ pattern and COD values over a representative time lapse for one of the experimental systems here considered. *Note*: times when re-fillings reached the system´s influent are shown as vertical light grey dot dashed lines; alarm signals are shown as dark grey dashed lines with end arrows.

Vertical light-grey dot dashed lines represent times when re-fillings reached the system´s influent, resulting in a rapid increase in organic matter concentration, which is in turn represented by the black dots depicting sampled COD values. COD values dropped as low as 60 mg $O_2$/L right before the refilling and leaped up to values as high as 378 mg $O_2$/L. The rapid increase of COD at the beginning of each cycle was immediately followed by a steep increase of $E_{cell}$ (up to ca. 400 mV) in most of the cases during the study period. It can also be seen, that $E_{cell}$ generally followed the steadily decrease of COD concentration (down to ca. 100 mV), but with a variable delay. Although there is a visual correlation between $E_{cell}$ and COD concentrations, the authors were not able to find a clear correlation between COD and the respective $E_{cell}$ values, as shown in Figure S6. This was probably due to the fact that COD decrease was followed by a $E_{cell}$ decrease but with a variable time delay.

Results obtained showed that the bioindication range decreased with system age or, more precisely, due to the accumulated organic matter (Figure S7). A good bioindication range could be achieved for systems aged ca. 3 weeks (after start-up and biofilm establishment time), for around 3 to 4 weeks (up to an accumulated organic loading of ca. 200 g COD/$m^2$), after that the average $E_{cell}$ signal range decreased from ca. 100-350 mV to ca. 250-350 mV. The decreased lower limit of $E_{cell}$ potential possibly reflects the effects of accumulated organic matter and endogenous respiration on voltage generation after longer operation time.

In light of these results, CW-MFC could be used as a qualitative tool to track sudden increases of COD at the influent of a wetland. The proposed alarm system to track a rapid influent COD increase is based on the steepness and the duration of a positive slope of $E_{cell}$, as well as the condition that no alarm signal was raised within less than 2 hours. The parameters Minimum Sum of Slopes (MSS), Minimum Slope Limit (MSL) and Minimum Alarm Interval (MAI) can be changed in order to calibrate the alarm tool; e.g. by decreasing MSS in order to also capture shorter lasting increases or by decreasing the MSL in order to also raise an alarm for a not so steep increases (it will take longer to reach MSS though). The datalogger recording interval (RI) is set to 15 minutes. Equation 1 shows the calculation of the slope by subtracting the slope ($S_i$) of the last reading from the present reading. Equation 2 sums up $S_i$ as long as $S_i$ is positive (i.e. increasing), as soon as it is negative (i.e. decreasing) it is reset to 0. Equation 3 calculates the time since the last alarm was raised. Finally, Equation 4 describes the conditions needed in order for the alarm system to be triggered; (1) $S_i$ has to be higher then the MAI, ensuring that the slope is steep enough, (2) $SS_i$ has to be higher than MSS, ensuring that the slope is increasing over a longer time, and (3) $AI_i$ has to be higher than MAI, ensuring that the last alarm is longer than 2 hours ago. If any of these three conditions is not met, no alarm is raised.

Below is the description of parameters and equations used for the calculation of the alarm signal.

Variable (calibration) parameters included in the "alarm-tool" are the following:

*MSS (Minimum Sum of Slopes)* = 140 *mV/h*

*MSL (Minimum Slope Limit)* = 16 *mV/h*

*MAI (Minimum Alarm Interval)* = 2 *hours*

Non-variable parameter included in the "alarm-tool" is the following:

$$RI\ (Recording\ Interval) = t_i - t_{i-1} = 0.25\ hours$$

The "alarm-tool" raises based on the following equations (Eq.1 to Eq.4):

Eq. 1

$$Si\ (Slope) = \frac{(Ecell_i - Ecell_{i-1})}{RI}\ i \in \mathbb{N}^+$$

Eq. 2

$$SSi \ (Sum \ of \ Slopes) = \sum_{j=a+1}^{i} s_j : \begin{cases} s_a \leq 0 \ a \in \ \mathbb{N}^+ \\ sj > 0, \forall j \in [a+1, i] \ j \in \ \mathbb{N}^+ \end{cases}$$

Eq. 3

$$AI_i \ (Alarm \ Interval) = RI \ \times (i - b) : \begin{cases} Alarm_b = 1 \ b \in \ \mathbb{N}^+ \\ Alarm_c = 0, \forall c \in [b, i] \ c \in \ \mathbb{N}^+ \end{cases}$$

Eq.4

$$A_i \ (Alarm)(S, SS, AI) = \begin{cases} 0 : (S_i \leq SL \ \lor \ SS_i \leq MSS \lor AI_i \leq \ MAI) \\ 1 : (S_i > SL \ \land \ SS_i > MSS \land AI_i > \ MAI) \end{cases}$$

The above presented alarm system was able to track most (75-80% across the three systems) of the episodes of rapid COD increase along the study period. Figure 6 shows the alarm signal at the times it was raised, resulting in an alarm response after around 2-4 hours in respect to the new wastewater reaching the influent.

## 3.8.-Final remarks

Our results suggest that the bioindication range decreased over time with the lower limit of $E_{cell}$ voltage increasing while the upper $E_{cell}$ voltage stayed constant. One important aspect that was observed as well is that the increase in organic matter concentration is followed immediately by the $E_{cell}$ potential whereas the decrease in organic matter concentration is only followed by $E_{cell}$ after a certain variable delay. The decrease in bioindication range as well as the delayed decrease in $E_{cell}$ compared to organic matter concentration could both be due to the above described generation of background level current due to oxidation of accumulated solids and through endogenous metabolism. Therefore, it is difficult to quantify organic matter concentration and water quality by means of CW-MFCs. However, the fast and reliable response to increasing organic matter loading open a feasible scenario on the use of CW-MFC as qualitative alarm-like assessment tool to adequately track a sudden increase in organic matter concentration.

### Conclusions

Positive linear relationships could be established between $E_{cell}$ generated and inlet COD concentration. This linearity was found both for gravel and graphite MFCs. Although, the best results were obtained after 20h of contact time, our biosensors showed correlations with R-squared values higher than 0.8 also for contact times higher than 10 hours, regardless the material used.

Graphite based MFCs showed better linear correlations than gravel systems regarding the COD range here tested.

Abiotic current component accounted for less than 2% of biologic current regardless the anodic material considered.

Organic matter contained in wastewater is not the only component of the electric current generated by MFC when implemented in CW. Organics trapped within the system and endogenous respiration could be also contributing the the current generated, therefore, compromising the long term accuracy of the tool.

CW-MFCs implemented in a pilot plant showed that a rapid increase in organic matter concentration was rapidly followed by an increase in CW-MFC cell potential, leading the authors to the conclusion that CW-MFC could be used as a dual-response "alarm-tool" for qualitative continuous influent water quality assessment.

## Acknowledgments

Authors thank Noelia Gómez,Claudia Jimenez, Marta Fernández, Fabio Muccioli and Ivan Genovese for their help, as undergraduate and graduate students, during the construction and experimental periods. Clara Corbella kindly acknowledges her PhD scholarship (2014 FI_AGAUR, Generalitat de Catalunya). Marco Harlt acknowledges the European Union´s Horizon 2020 research and innovation programme under the Marie Skłodowska-Curie grant agreement No 676070. This communication reflects only the authors' view and the Research Executive Agency of the EU is not responsible for any use that may be made of the information it contains.

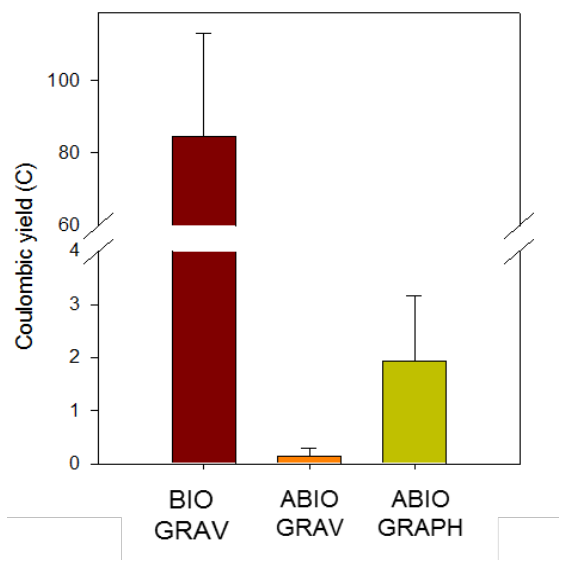

**Figure S1.** Coulombic yield for MFC operated under abiotic and biotic conditions.

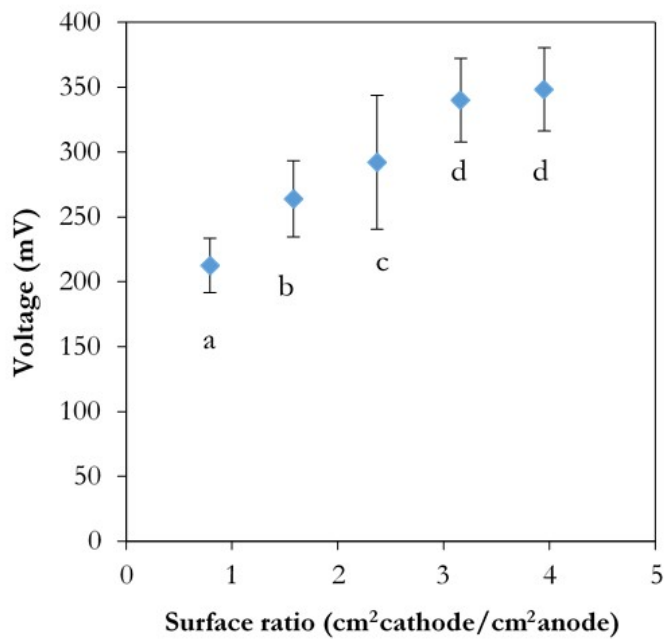

**Figure S2.** Voltage recorded as function of the cathode to anode surface ratio. *Note*: letters are used to indicate groups of statistical difference.

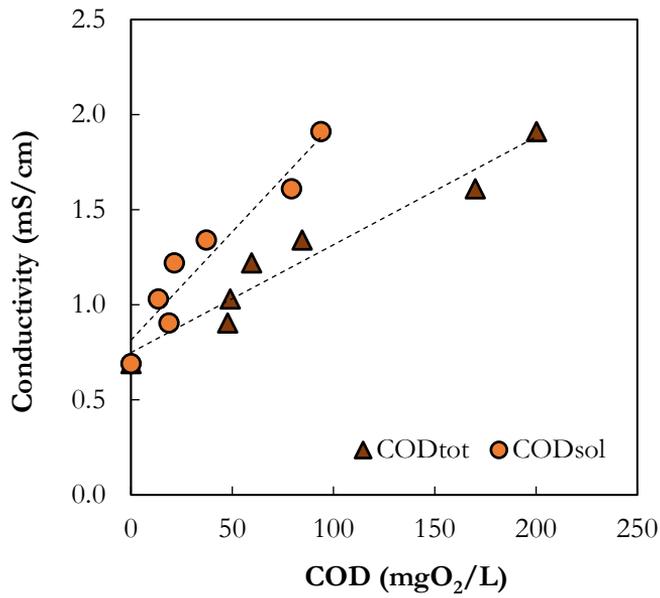

**Figure S3.** Linear relationship between total (triangles) and soluble (circles) chemical oxygen demand (COD) and wastewater conductivity. *Note*: R-squared values were of 0.94 (total COD) and 0.92 (soluble COD).

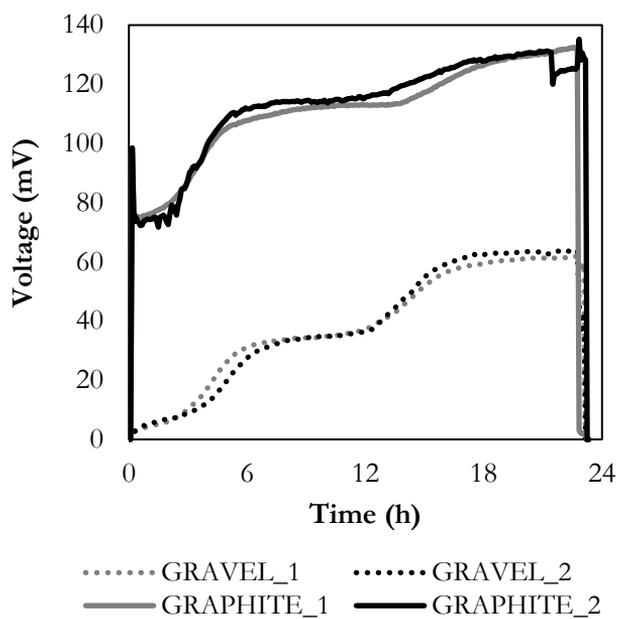

**Figure S4.** Representative pattern for gravel and graphite-based MFC during bioindication experiment.

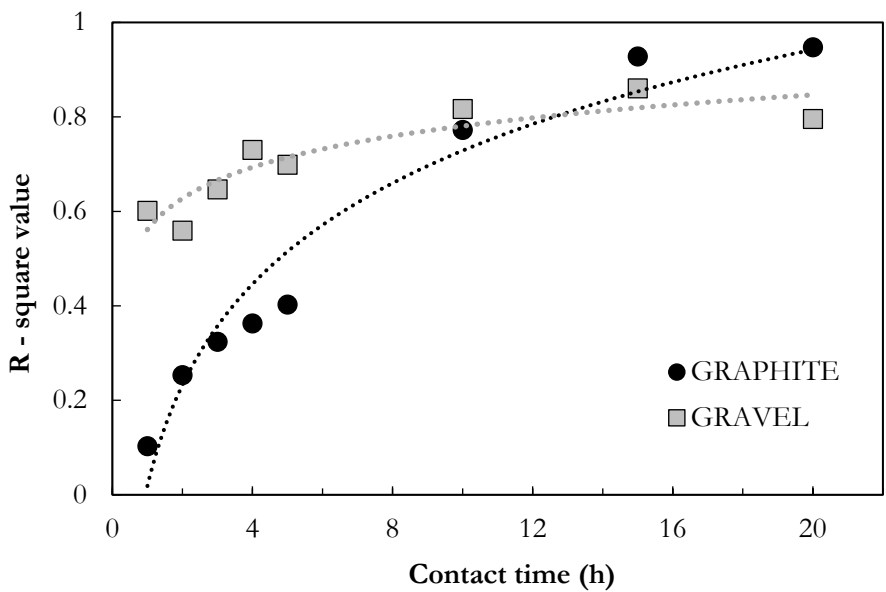

**Figure S5.** R-squared values obtained from the linear relationships stablished between the current generated and the COD tested as function as the contact time (h) during E1.

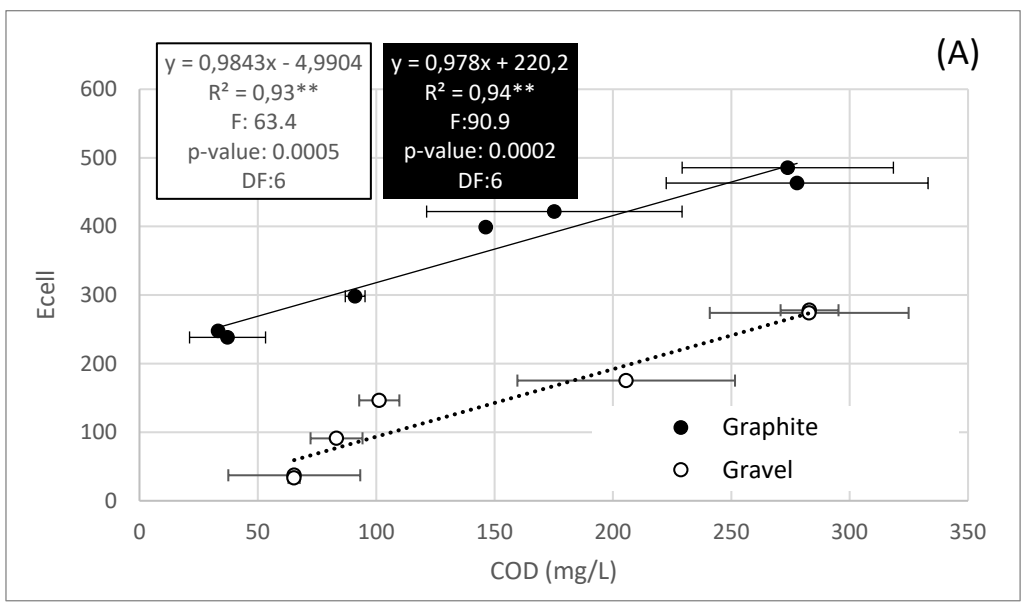

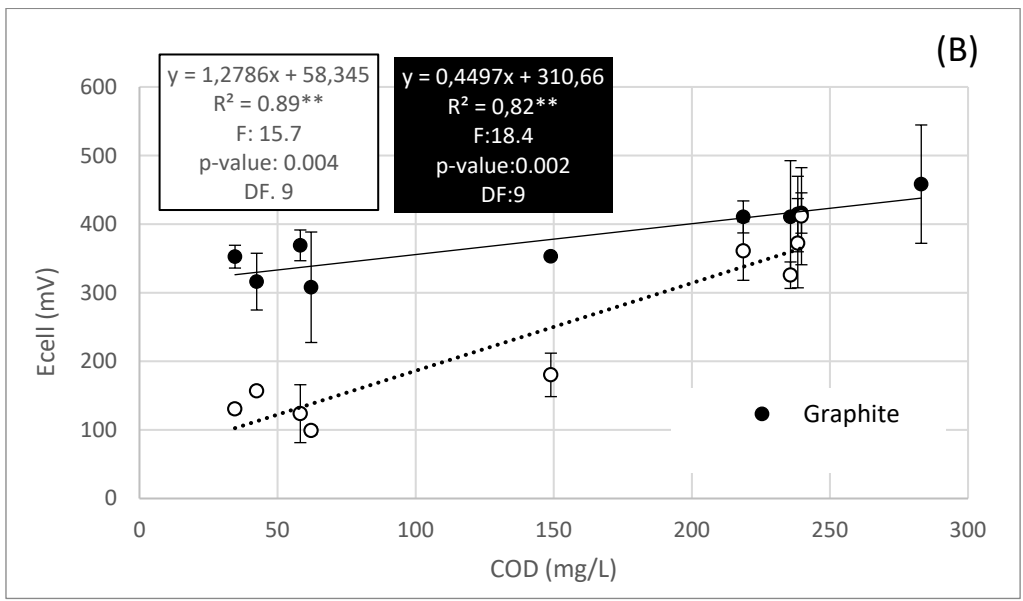

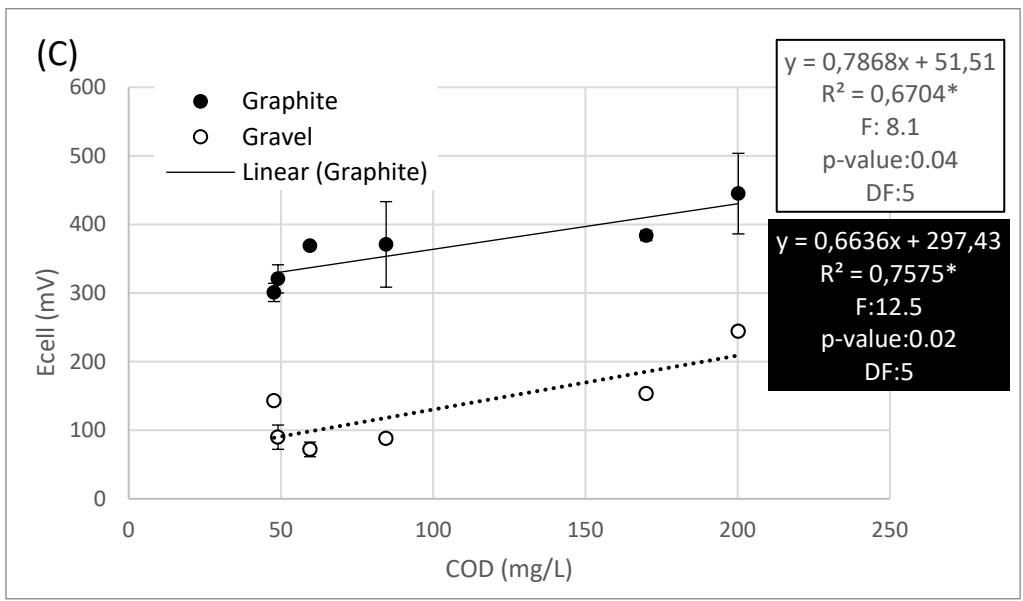

**Figure S6.** Linear regression for each batch performed. (A) Batch 2; (B) Batch 3; (C) Batch 4. Note: Asterisk represents the level of statistical significance of the linear regression (*p-value<0.05; **p-value<0.01).

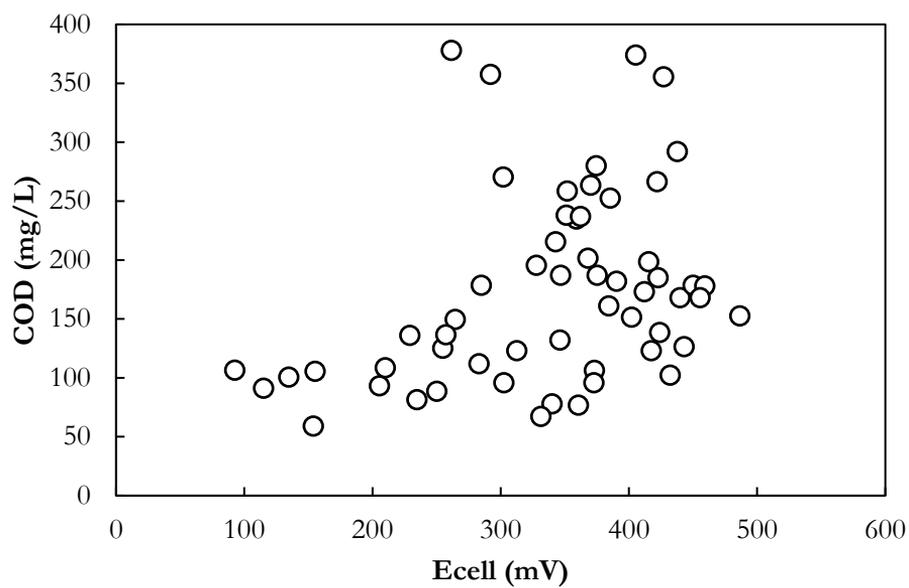

**Figure S7.** Correlation between E<sub>cell</sub> and COD concentration for one of the two newer systems (n=57)

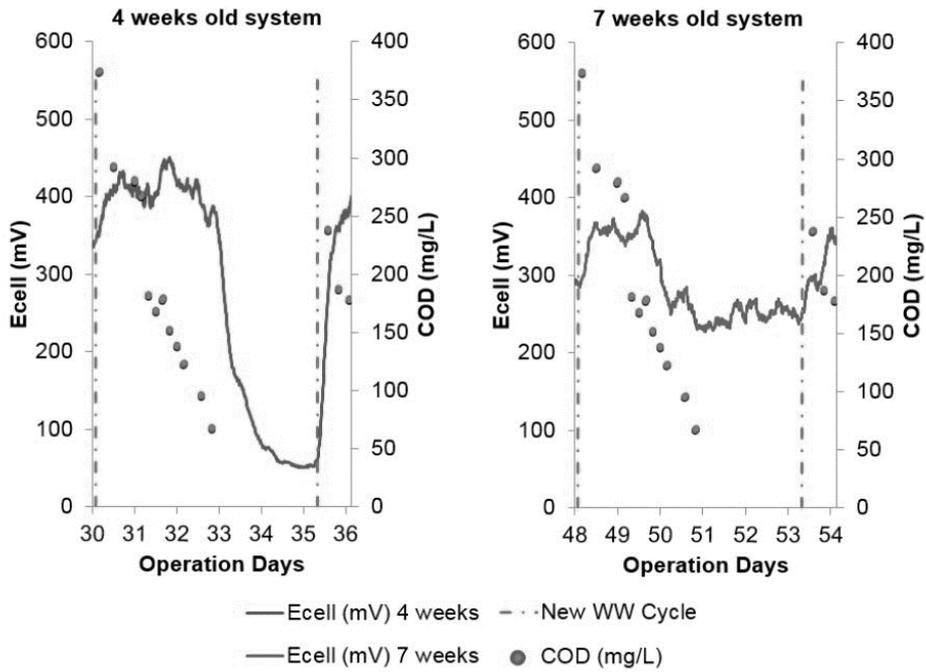

**Figure S8.** E<sub>cell</sub> pattern and COD values for 4 weeks (left) and 7 weeks old (right) systems (the x-axis shows respective system operation days). *Note*: times when re-fillings reached the system´s influent are shown as vertical orange dot dashed lines.